\documentclass[aps,amsmath,amssymb,nofootinbib,twocolumn,pr]{revtex4-1}
\usepackage{graphicx}
\usepackage{color}
\usepackage{verbatim} 
\usepackage{color}
\usepackage{bbold}
\usepackage{dcolumn}
\usepackage{bm}
\usepackage{upgreek}
\usepackage[unicode=true,colorlinks=true,citecolor=blue,urlcolor =blue]{hyperref}
\newcommand*\xoverline[2][0.75]
\begin{document}

\title{Exciton oscillator strength in two-dimensional Dirac materials}

\author{N.~V.~Leppenen}
\author{L.~E.~Golub} 	
\author{E.~L.~Ivchenko} 	
\affiliation{Ioffe Institute, 194021 St. Petersburg, Russia}


\begin{abstract}
Exciton problem is solved in the two-dimensional Dirac model with allowance for strong electron-hole attraction. The exciton binding energy is assumed smaller than but comparable to the band gap.
The exciton wavefunction is found in the momentum space as a superposition of all four two-particle states including electron and hole states with both positive and negative energies. The matrix element of exciton generation is shown to depend on the additional components of the exciton wavefunction. Both the Coulomb and the Rytova--Keldysh potentials are considered. The dependence of the binding energy on the coupling constant is analyzed for  the ground and first excited exciton states. The binding energy and the oscillator strength are studied as functions of the environmental-dependent dielectric constant for real transition metal dichalcogenide monolayers. We demonstrate that the multicomponent nature of the exciton wavefunction is crucial for description of resonant optical properties of two-dimensional Dirac systems.
\end{abstract}

\maketitle

\section{Introduction}

In recent decade, a new family of condensed-matter systems is being investigated which is classified as Dirac materials~\cite{DM_review}.  
The main feature of the Dirac materials is a moderate value of the band gap in comparison to other energy scales. These systems can be effectively described by the Dirac equation. There are some specific properties of two-dimensional (2D) Dirac materials, e.g. a half-integer Chern number~\cite{beyond_gr}.
A prominent example of two-dimensional massive Dirac materials is 
transition metal dichalcogenide (TMD) monolayers. They are extremely attractive due to strong Coulomb effects which are probed by various optical spectroscopy methods where a series of strong exciton and trion resonances are present~\cite{rev1,rev2,GlazovChernikov}.
Exciton resonances as well as continuous absorption spectra are very different from those in  conventional semiconductors. In particular, the exciton binding energy is comparable with the band gap. 

Strong Coulomb interaction in the Dirac materials makes invalid the traditional theoretical approach to the exciton problem based on the parabolic band approximation.
Hence, it is insufficient to consider the exciton problem assuming the Coulomb interaction as a small perturbation as it has been done in Refs.~\cite{Silin,Weak_int}.
Furthermore, the Coulomb scattering in the Dirac systems involves both intra- and interband processes~\cite{Scattering_gapped_graphene}.
In fact, the exciton state becomes a superposition of two-particle excitations with both the conduction and valence band single-particle states: all the four possibilities are realized with the electron and the hole having both signs of energy~\cite{Two_body_graphene}.
Therefore, ignoring the negative-energy electron states and positive-energy hole states used in a number of works, see e.g. Refs.~\cite{Exc_top_ins,MacDonald_2015,Trushin_2016,Trushin_2018}, is inappropriate. Indeed, the inter- and intraband Coulomb energies are of the same order in the Dirac materials because a parameter making them strongly different in ordinary semiconductors is a ratio of the exciton Rydberg energy to the band gap. In Ref.~\cite{Somm_fact_TMD} the exciton problem in the 2D Dirac materials was reduced, without justification, to an analytically solvable system of two equations equivalent to the problem of a charged particle bound to an immobile Coulomb center. The nonequivalence of the bound-particle and motionless exciton problems is a specific feature of the non-parabolic energy spectrum of free electrons and holes in the Dirac materials where the two-particle Schr\"odinger equation cannot be reduced  to a single-particle one. Because of the unjustified approach, both the exciton level positions and the Sommerfeld factor calculated in the work \cite{Somm_fact_TMD}  are questionable.

A correct approach has been used in Refs.~\cite{Exc_trion_TMD,Exc_TMD_energies} where numerical solutions of four coupled differential equations for the exciton wavefunction components have been obtained and exciton energies have been calculated. However, the exciton oscillator strength calculation performed in Ref.~\cite{Exc_trion_TMD} ignores the  four-component form of the wavefunction.

In this work, the theory of excitons in the 2D Dirac materials is developed accounting for the exciton binding energy being comparable (but smaller) than the band gap and the exciton oscillator strength is calculated as a function of the electron-hole coupling strength.

The paper is organized as follows. In Sec.~\ref{exciton} we present equations for the four-component exciton wavefunction and derive a general expression for the oscillator strength.
In Sec.~\ref{bound} we calculate and discuss the binding energy and oscillator strength for both the 2D Coulomb and Rytova--Keldysh potentials.  
Concluding remarks are  presented in Sec.~\ref{Conc}.

\section{Exciton in the 2D Dirac model}
\label{exciton}
We consider here the 2D Hamiltonian describing the behaviour of electrons in the two valleys $K$ and $K'$ related by the time inversion operation ${\cal T}$. 
The single-electron effective Hamiltonian in the $K$ valley has the form
\begin{equation} \label{HamiltK}
{\cal H}^K(\bm k)= \hbar v_0 \bm \sigma \cdot \bm k + {E_g\over 2}\sigma_z\:,
\end{equation}
where $\bm k=(k_x,k_y)$ is the 2D wavevector counted from the $\bm K$ point of the 2D Brillouin zone, $\sigma_{x,y,z}$ are the pseudospin Pauli matrices acting in the basis of two Bloch functions 
$\psi^K_1, \psi^K_2$  at the $K$ point ({$\bm k= 0$}), 
and $v_0$, $E_g$ are the Dirac velocity  and the energy gap. The eigenenergies are given by
\begin{equation} \label{relativ}
\epsilon_{\lambda \bm k} = \lambda \epsilon_k,
\quad \lambda = \pm, \quad  \epsilon_k = \sqrt{(E_g/2)^2+(\hbar v_0k)^2}\:.
\end{equation}
The corresponding eigenfunctions {can be written as a sum of two products of envelopes depending on ${\bm k}$ and the Bloch functions at ${\bm k} = 0$:
\begin{equation} \label{psiKk}
\Psi^K_{\lambda,{\bm k}}({\bm \rho}) = \psi^K_{\lambda,{\bm k},1}({\bm \rho}) \psi^K_1 + \psi^K_{\lambda,{\bm k},2}({\bm \rho}) \psi^K_2 \:,
\end{equation}
where ${\bm \rho} = (x,y)$ is the 2D radius-vector.
According to Eq.~(\ref{HamiltK}) one can conveniently present the envelope functions as two-component spinors 
\begin{equation} \label{envelope}
\psi^K_{\lambda,{\bm k}}({\bm \rho}) = {\rm e}^{{\rm i} {\bm k}\cdot{\bm \rho}} u^K_{\lambda,{\bm k}}\:,
\end{equation}
where 
$u^K_{\lambda,{\bm k}}$ are eigencolumns of the Hamiltonian (\ref{HamiltK})}
\begin{equation}
u_{+,\bm k}= 
\begin{bmatrix}
T_+ \text{e}^{-i\varphi_{\bm k}/2} \\
T_- \text{e}^{i\varphi_{\bm k}/2} 
\end{bmatrix},
\quad
u_{-,\bm k}= 
\begin{bmatrix}
-T_- \text{e}^{-i\varphi_{\bm k}/2} \\
T_+ \text{e}^{i\varphi_{\bm k}/2} 
\end{bmatrix}.
\end{equation}
Here $\varphi_{\bm k}$ is the azimuth angle of the vector $\bm k$, and 
\begin{equation} \label{Tcoef}
T_\pm = \sqrt{\frac12 \left( 1 \pm \frac{E_g}{2\epsilon_k} \right)} \:.
\end{equation} 

The exciton is a two-particle electron-hole state. For definiteness, we consider excitons formed by an electron belonging to the $K$ valley and a $K'$ valley hole representing the missing electron also in the $K$ valley. The exciton wavefunction satisfies the Schr\"odinger equation \cite{Weak_int,Exc_TMD_energies}
\begin{eqnarray} \label{HeHh}
\left[ {\cal H}^K(\hat{\bm k}_e) \otimes \mathbb{1}  +  \mathbb{1} \otimes {\cal H}^{h,K'}(\hat{\bm k}_h)  + V({\bm \rho})\right]  \Psi_{\rm exc}({\bm \rho}_e, {\bm \rho}_h) \nonumber\\ = E \Psi_{\rm exc}({\bm \rho}_e, {\bm \rho}_h)\:, \hspace{3 cm}
\end{eqnarray}
where ${\bm \rho}$ is the difference ${\bm \rho}_e - {\bm \rho}_h$, 
$V({\bm \rho})$ is the attractive (negative) Coulomb potential,
$ \Psi_{\rm exc}({\bm \rho}_e, {\bm \rho}_h)$ is a column consisting of four components $(\psi_{++},\psi_{+-},\psi_{-+},\psi_{--})$ dependent on the electron and hole coordinates, respectively $x_e,y_e$ and $x_h,y_h$, $\hat{\bm k}$ is the differential operator $-{\rm i} {\bm \nabla}$, ${\cal H}^K(\hat{\bm k}_e)$ and ${\cal H}^{h,K'}(\hat{\bm k}_h)$ are the electron and hole effective Hamiltonians, and we use the index notation $++,+-,-+,--$ instead of $AA,AB,BA,BB$ \cite{Two_body_graphene} or $cc, cv, vc, vv$ \cite{Weak_int,Exc_TMD_energies}. 

\subsection{Relation between the $K$ and $K'$ valley states} The states in the $K'$ valley are related with those in the $K$ valley by the time reversal  operator 
\begin{equation} \label{calT}
{\cal T}=-i\sigma_2 {\cal K}_0\:,
\end{equation}
with ${\cal K}_0$ being the complex conjugate operation and $\sigma_2$ being the second spin Pauli matrix. Particularly, the energy spectrum in the  $K'$ valley is also described by Eq.~(\ref{HamiltK}) and the sign $\lambda$ has the same meaning. Moreover, there is a linear relation between the Bloch wave functions $\psi^{K'}_j \:(j= 1,2)$ at the $K'$ point and the functions ${\cal T}\psi^K_{j'} \:(j' = 1,2)$. We take this relation in the form
\begin{equation} \label{K'K}
\psi^{K'}_1= - {\cal T} \psi^K_2 \:, \qquad \psi^{K'}_2=  {\cal T} \psi^K_1 \:.
\end{equation}
In this case the indices $j, j'$ can be conceived as the spin components $\pm 1/2$. In the chosen basis
the effective Hamiltonian in the $K'$ valley reads 
\begin{equation} \label{both}
{\cal H}^{K'}({\bm k}) = \hbar v_0 \left( - \sigma_x k_x + \sigma_y k_y \right) - {E_g\over 2}\sigma_z\:.
\end{equation}
For simplicity, we omit in Eq.~(\ref{both}) rigid band shifts due to the spin-orbit interaction.

The single hole states are defined as the empty electron states as follows: the missing electron state $|e,K, -\lambda, -\bm k \rangle$ with the energy $\epsilon_{-  \lambda,-{\bm k}} = -\lambda \epsilon_{k}$ in the $K$ valley is equivalently described by the hole state $|h, K', \lambda, {\bm k} \rangle$ with the energy $\lambda \epsilon_k$ in the opposite $K'$ valley. In the symbolic form the relation between the electron and hole representations can be written in the following way
\begin{eqnarray} \label{symbolic}
| e,K, -\lambda_h,- {\bm k}_h \rangle = {\cal T} | h, K', \lambda_h, {\bm k}_h \rangle\:. \nonumber
\end{eqnarray}
For the relation (\ref{K'K}) the hole effective Hamiltonian is expressed via ${\cal H}^K({\bm k})$ as
\begin{eqnarray} \label{Hamilth}
{\cal H}^{h,K'}({\bm k}) = - {\cal H}^{K'}({\bm k}) = {\cal H}^K(-{\bm k})\:.
\end{eqnarray}

\subsection{The exciton wavefunction} 
In this work we take the electron-hole total momentum $\hbar({\bm k}_e + {\bm k}_h)$ to be zero which allows us to set ${\bm k}_e = - {\bm k}_h \equiv {\bm k}$ and to seek the exciton wavefuncion dependent on 
$\bm \rho$. 
In this case the exciton wavefunction expansion in the states of noninteracting electron-hole pairs $|e, \lambda_e, {\bm k}_e; h, \lambda_h, {\bm k}_h \rangle$ is written as follows
\begin{equation} \label{Clelh}
\vert {\rm exc} \rangle = \sum_{\lambda_e \lambda_h}\sum_{\bm k} C_{\lambda_e\lambda_h}(\bm k) |e, \lambda_e, {\bm k}; h, \lambda_h, -{\bm k} \rangle \:,
\end{equation}
where $C_{\lambda_e\lambda_h}$ are the expansion coefficients dependent on the ${\bm k}$ vector.
In what follows, in order to simplify the normalization procedure, we set the sample area to unity. The expansion coefficients satisfy a set of four coupled equations 
\begin{align} 
\label{C_syst}
& 
(\lambda_e + \lambda_h)\epsilon_k   C_{\lambda_e \lambda_h}({\bm k}) \\ 
&+
\sum\limits_{\lambda'_e \lambda'_h}\sum_{{\bm k}'} {\cal J}_{\lambda_e \lambda_h; \lambda'_e \lambda'_h}({\bm k} \leftarrow {\bm k}')   C_{\lambda'_e \lambda'_h}({\bm k}') \nonumber 
= E   C_{\lambda_e \lambda_h}({\bm k})\:,
\end{align}
where the Coulomb scattering matrix element is formally given by
\begin{eqnarray} \label{scattexc}
&&{\cal J}_{\lambda_e \lambda_h; \lambda'_e \lambda'_h}({\bm k} \leftarrow
{\bm k}') \\ &&= \langle e, \lambda_e, {\bm k}; h, \lambda_h, -{\bm k} | V | e, \lambda'_e, {\bm k}'; h, \lambda'_h, -{\bm k}' \rangle \:. \nonumber
\end{eqnarray} 
Bearing in mind the relation between the hole state and the missing electron state we can present (\ref{scattexc}) as
\begin{eqnarray} \label{scattexc2}
&&  V(\bm q) \int \left[ \psi^K_{\lambda_e,{\bm k}} ({\bm \rho})\right]^{\dag}
{\rm e}^{{\rm i} {\bm q}{\bm \rho}} \psi^K_{\lambda'_e,{\bm k}'} ({\bm \rho}) d {\bm \rho} \\&& \times \int \left[ \psi^K_{-\lambda'_h,  {\bm k}'} ({\bm \rho}')\right]^{\dag}{\rm e}^{-{\rm i} {\bm q}{\bm \rho}'} \psi^K_{-\lambda_h,{\bm k}} ({\bm \rho}') d {\bm \rho}' \:,\nonumber   
\end{eqnarray} 
where ${\bm q} = {\bm k} - {\bm k}'$ and $V({\bm q})$ is the 2D Fourier-image of the potential $V(\bm \rho)$. Substituting (\ref{envelope}) into the integrands we obtain
instead of Eq.~(\ref{scattexc2})
\begin{align} \label{scattexc3}
 V({\bm q}) \left( u^{\dag}_{\lambda_e, {\bm k}} u_{\lambda'_e, {\bm k}'} \right) \left( u^{\dag}_{-\lambda'_h, -{\bm k}'} u_{-\lambda_h, -{\bm k}} \right)\:.
\end{align}
By using the identity
\begin{equation} \label{identity1}
u^{\dag}_{\lambda', {\bm k}'} u_{\lambda, {\bm k}}  = \lambda \lambda' u^{\dag}_{-\lambda, {\bm k}} u_{-\lambda', {\bm k}'}
\end{equation}
we can rewrite  the last term in Eq.~(\ref{scattexc3}) as 
\[
\lambda_h \lambda'_h\ u^{\dag}_{\lambda_h, {\bm k}} u_{\lambda'_h, {\bm k}'} \:.
\]
It is convenient to introduce the coefficients
\begin{equation} \label{coefficients}
{\cal C}_{\lambda_e \lambda_h}({\bm k})=\lambda_h C_{\lambda_e\lambda_h}({\bm k})\:.
\end{equation}
The set of equations for ${\cal C}_{\lambda_e \lambda_h}({\bm k})$ coincides with the set~(\ref{C_syst}) where  the scattering matrix element has the form
\begin{align} \label{scattexc4}
&{\cal J}_{\lambda_e \lambda_h; \lambda'_e \lambda'_h}({\bm k} \leftarrow {\bm k}') \nonumber \\
&= V({\bm q}) \left( u^{\dag}_{\lambda_e, {\bm k}} u_{\lambda'_e, {\bm k}'} \right) \left( u^{\dag}_{\lambda_h, {\bm k}} u_{\lambda'_h, {\bm k}'} \right)\:. \nonumber
\end{align}
We note that the exciton is formed by the free electron-hole pair states with a sum excitation energy $(\lambda_e + \lambda_h)\epsilon_k$ which takes not only values $2\epsilon_k$ but also zero and  $-2\epsilon_k$. This accounts for both intra- and inter-band scattering of free carriers by the potential $V$.

\subsection{Matrix elements of exciton optical generation}
We take the electron-photon interaction in the form
$V_{{\rm e}\mbox{-}{\rm ph}} = - c^{-1}  \int  j_{\mu}({\bm \rho})A_{\mu}({\bm \rho},t) d {\bm \rho}$, where ${\bm A}({\bm \rho},t)$ is the vector-potential of the plane electromagnetic wave of the frequency $\omega$, and $ {\bm j}({\bm \rho})$ is the operator of the electric current density. Then the exciton excitation matrix element can be written as 
\begin{equation} \label{el-phot}
\hspace{1.5 cm}\langle {\rm exc} | V_{{\rm e}\mbox{-}{\rm ph}} |0  \rangle = - \frac{A}{c}
{\rm e}^{-{\rm i} \omega t} M({\bm e})\:,
\end{equation}
where $M$ is the current density matrix element
\begin{equation} \label{Current}
 M({\bm e}) = \sum_{\lambda_e \lambda_h {\bm k}} \lambda_h{\cal C}^*_{\lambda_e \lambda_h}({\bm k}) 
\langle \lambda_e, {\bm k}; \lambda_h, -{\bm k} | {\bm e} \cdot {\bm j}(0) | 0 \rangle\:,
\end{equation}
$A, {\bm e}$ are the amplitude and the polarization unit vector of the electromagnetic wave, and ${\bm j}(0)$ is the Fourier component of the electron current density operator taken at zero wavevector. The matrix element of the electron-hole pair excitation is written in the electron representation as
\begin{eqnarray}
&&\langle e,\lambda_e, {\bm k}; h,\lambda_h, -{\bm k} | {\bm e}\cdot {\bm j}(0) | 0 \rangle \nonumber \\ &&=  e\ \langle e, K, \lambda_e, {\bm k} | \left( {\bm e}\cdot{\bm v} \right) {\cal T}| h, K',\lambda_h, -{\bm k} \rangle\: \nonumber
\\ &&= e \left( u^K_{\lambda_e, {\bm k}} \right)^{\dag} {\bm e}\cdot{\bm v}\ u^K_{-\lambda_h,{\bm k}}\:, \nonumber
\end{eqnarray}
where the velocity operator
\[
{\bm v} = \frac{1}{\hbar} \frac{\partial {\cal H}^K({\bm k})}{\partial {\bm k}} = v_0 {\bm \sigma}\:.
\]
As a result, we obtain instead of Eq.~(\ref{Current})
\begin{equation} \label{el-phot2}
M({\bm e}) = e v_0 \sum_{\lambda_e \lambda_h {\bm k}}\lambda_h {\cal C}^*_{\lambda_e \lambda_h}({\bm k}) \left( u^K_{\lambda_e,{\bm k}} \right)^{\dag}  {\bm e}\cdot {\bm \sigma}\ u^K_{-\lambda_h, {\bm k}}\:. 
\end{equation}
We remind that for the right and left circular polarizations the unit vector ${\bm e}$ reads
\[
{\bm e}_{\sigma^+} = \frac{\hat{e}_x + {\rm i} \hat{e}_y}{\sqrt{2}}\:,
\quad {\bm e}_{\sigma^-} = \frac{\hat{e}_x - {\rm i} \hat{e}_y}{\sqrt{2}}\:,
\]
where $\hat{e}_x$ and $\hat{e}_y$ are the unit vectors pointing in the directions $x$ and $y$.

So far as we know, it is the first time when the expression for  the exciton optical matrix element contains all the four terms rather than only one term with $\lambda_e=+$ and $\lambda_h = +$. The following calculation shows that the additional terms remarkably contribute to the exciton oscillator strength if the exciton binding energy is not very small as compared to the band gap.
\subsection{Solution to the exciton wavefunction}
While solving the two-body problem in graphene in the real space, 
Sabio et al.~\cite{Two_body_graphene} noticed that the problem of four-component two-particle wave function $\Psi_{j'j}({\bm \rho}_1, {\bm \rho}_2) ~(j', j = A, B)$, for zero total center-of-mass momentum, is decoupled under a certain unitary transformation into a set of equations for three transformed components and an independent equation for the remaining component. The similar property holds also for the Fourier coefficients ${\cal C}_{\lambda_e, \lambda_h}$. 
Under the unitary transformation of the two components
\begin{equation}
{\cal C}^\pm={{\cal C}_{+-}\pm{\cal C}_{-+} \over \sqrt{2}},
\end{equation}
the equation set  for ${\cal C}_{\lambda_e, \lambda_h}$ is split off into a single equation for ${\cal C}^-$ and a reduced system of three interconnected equations for ${\cal C}_{++}$, ${\cal C}_{--}$ and ${\cal C}^{+}$. We define a three-component vector $\bm{\mathcal C}(\bm k)$ with the components ${\cal C}_{++}(\bm k), {\cal C}^{+}(\bm k), {\cal C}_{--}(\bm k)$ satisfying the equation
\begin{align} \label{syst_3_eq}
& E\bm{\mathcal C}(\bm k) = \bm{\mathcal H}_0(k)\bm{\mathcal C}(\bm k)\\
&+\sum_{\bm k'}V(|\bm k-\bm k'|)\sum_{l=0, \pm 1} {\bm F}_l(k,k')\text{e}^{il(\varphi_{{\bm k}'}-\varphi_{\bm k})}\bm{\mathcal C}(\bm k')\:. \nonumber
\end{align}
Here $\bm{\mathcal H}_0(k)$ is the diagonal 3$\times$3 matrix
\begin{equation} \label{H0}
\left[ \begin{array}{ccc} 2\epsilon_k & 0 & 0 \\ 0& 0& 0 \\ 0 & 0 & -2\epsilon_k \end{array} \right]\:,
\end{equation}
and the 3$\times$3 matrix ${\bm F}_l(k,k')$ is a product of the 3$\times$1 matrix (a column) $S_l(k)$ and  the transposed matrix (a row) $S_l^T(k')$,  where
\begin{equation}
S_{\pm 1}(k)= \left[\begin{array}{c}T_\mp^2 \\ \pm\sqrt{2}T_+T_- \\ T_\pm^2\end{array}\right],
\quad
S_0(k)= \left[\begin{array}{c} \sqrt{2}T_+T_-  \\  T_+^2-T_-^2 \\ -\sqrt{2}T_+T_-\end{array}\right],
\end{equation}
and the coefficients $T_{\pm}(k)$ are introduced in Eq.~(\ref{Tcoef}).

From symmetry considerations of the studied two-valley band structure the motionless excitons should have a certain value of the angular momentum component~\cite{Exc_TMD_energies,Potemski}. This agrees with the kernel of Eq.~\eqref{syst_3_eq} depending on the phase difference $\varphi_{\bm k}-\varphi_{\bm k'}$, and we can seek the solutions in the form
\begin{equation} \label{angular}
\bm{\mathcal C} ({\bm k})= \bm{\mathcal C}_m(k) {\rm e}^{{\rm i} m \varphi_{\bm k}}\:,
\end{equation} 
where  $m = 0, \pm 1, \pm 2\dots$ 

Substituting ${\cal C}^+$ and ${\cal C}^-$ instead of ${\cal C}_{+-}$, ${\cal C}_{-+}$ in Eq.~(\ref{el-phot2}) we find that the coefficient ${\cal C}^-$ makes no contribution to the optical matrix element and obtain
\begin{equation} \label{osc_strength}
M({\bm e}) = e v_0 \sum\limits_{\bm k} \left( \text{e}^{{\rm i}\varphi_{\bm k}} e_- R_+ -  \text{e}^{-{\rm i}\varphi_{\bm k}} e_+ R_- \right)\:,
\end{equation}
where $e_{\pm} = e_x \pm {\rm i} e_y$ and
\begin{align}
\label{R}
&R_+({\bm k}) = T_+^2 {\cal C}^*_{++}(\bm k) + T_-^2 {\cal C}^*_{--}(\bm k) - \sqrt{2}T_+T_- {\cal C}^{+ *}(\bm k) \:, \nonumber\\
&R_-({\bm k}) = T_-^2 {\cal C}^*_{++}(\bm k) + T_+^2 {\cal C}^*_{--}(\bm k) + \sqrt{2}T_+T_- {\cal C}^{+ *}(\bm k) \:. 
\end{align} 
Particularly, it follows from here that for the circularly polarized light one has
\begin{equation} \label{osc_strength2}
M(\sigma^+) = \sqrt{2} e v_0 \sum\limits_{\bm k} \text{e}^{{\rm i}\varphi_{\bm k}} R_+({\bm k}) \:.
\end{equation}
We see that it is the exciton state with the angular harmonics $m=1$ which is optically active in the $\sigma^+$ polarization. 

\section{Results and Discussion} 
\label{bound}
We seek for the exciton eigenenergies and oscillator strength for two forms of the attractive electron-hole interaction relevant to the 2D Dirac materials and modeled  by (i)~the standard 2D Coulomb potential 
\begin{equation} \label{V_q} 
V_C(q) = -{2\pi e^2\over \varkappa q}
\end{equation}
and (ii) the Rytova--Keldysh potential \cite{Rytova,Keldysh}
\begin{equation} \label{V_q2} 
V_{RK}(q)= -{2\pi e^2\over \varkappa q(1+qr_0)}\:.
\end{equation}
Here $\varkappa$ is the half sum of the dielectric susceptibilities of 
materials surrounding the 2D layer,
and ${r_0=l/\varkappa}$ is the screening radius with the length $l$ determined by the susceptibility of the 2D layer~\cite{rev2}.

The two-body vector equation (\ref{syst_3_eq})  gives rise to bound (with $E < E_g$) and unbound (with $E > E_g$) excitons leading to discrete and continuous optical absorption. In the present work we focus our attention on the bound exciton states.

It should be noted that here we do not perform renormalization of the parameters $v_0$ and $E_g$ by the electron-electron interaction assuming they are already taken into account. This problem has been intensively studied in graphene~\cite{graphene_renorm}. An allowance for the renormalization for a finite band gap will be published elsewhere.

\subsection{Binding energy}

Let us start from the Coulomb potential with the Fourier image $V_C(q)$, Eq.~\eqref{V_q}. We study the dependence of the binding energy $E_g-E$ on the dimensionless interaction strength
\begin{equation} \label{ge2}
g = {e^2\over \varkappa \hbar v_0}.
\end{equation}

First of all we will analyze the equation~(\ref{syst_3_eq}) in the limit of small $g$ where the exciton state is formed by small values of $k$ so that we can set $T_+(k) \to 1$, ${T_-(k) \to 0}$, the matrices ${\bm F}$ become diagonal, ${F_{l; i'i} \to \delta_{i'i} \delta_{i,-l}}$, and the components ${\cal C}_{--}$, ${\cal C}^+$ vanish. The energy $2 \epsilon_k$ can be be written in the parabolic  approximation as ${E_g + \hbar^2 k^2/(2 \mu)}$, where $\mu$ is the exciton reduced mass $E_g/(4 v_0^2)$. Then, the equation for the remaining component ${\cal C}_{++}$ reduces to
\begin{eqnarray} \label{mass}
\left( E_g  + \frac{\hbar^2 k^2}{2 \mu} - E\right) \text{e}^{-{\rm i}\varphi_{\bm k}} {\cal C}_{++}({\bm k}) \hspace{1 cm} \\ + \sum\limits_{{\bm k}'} V(|{\bm k} - {\bm k}'|) \text{e}^{-{\rm i}\varphi_{{\bm k}'}} {\cal C}_{++}({\bm k}') = 0 \:. \nonumber
\end{eqnarray}
Thus, the exciton envelope function in the effective mass theory is related to ${\cal C}_{++}$ by
\begin{equation} \label{Psiexc}
\Psi_{\rm exc}({\bm k}) = \text{e}^{-{\rm i}\varphi_{{\bm k}}} {\cal C}_{++}({\bm k})\:.
\end{equation}
We see that the angular momentum component, $\ell$, of the exciton envelope is related with the integer $m$ in Eq.~(\ref{angular}) by $\ell = m -1$. Since we assume the Fermi velocity $v_0$ to be positive we can assign the angular momentum component $+1$ to the interband electron excitation in the $K$ valley. Therefore, the total $z$-component of the angular momentum equals $\ell + 1 = m$ and, for the $\sigma^+$ optical excitation, the optically allowed are the exciton states with $m = 1$. This selection rule agrees with Eq.~(\ref{osc_strength2}).

\begin{figure}[h]
\includegraphics[width=0.99\linewidth]{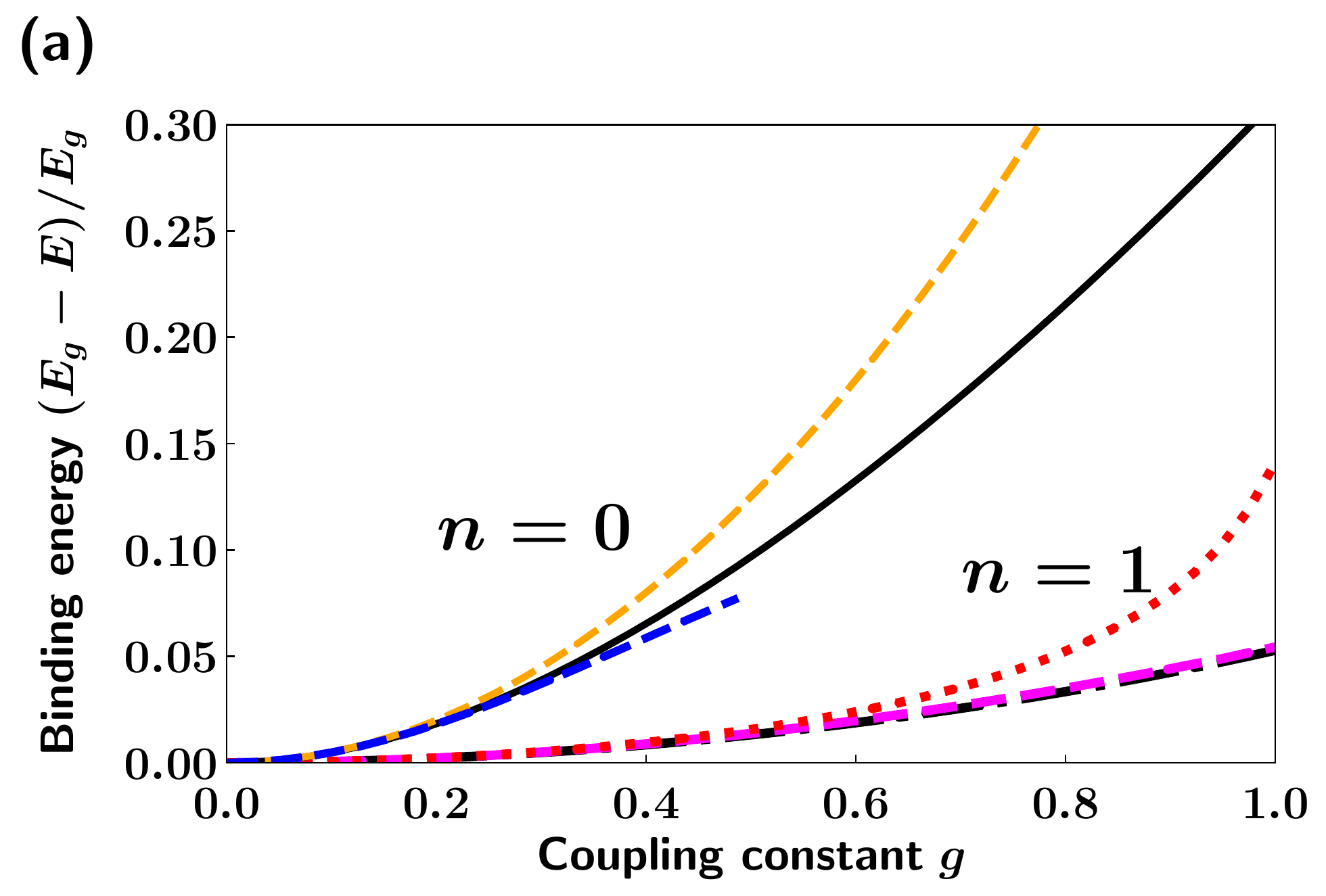}\\
\includegraphics[width=0.99\linewidth]{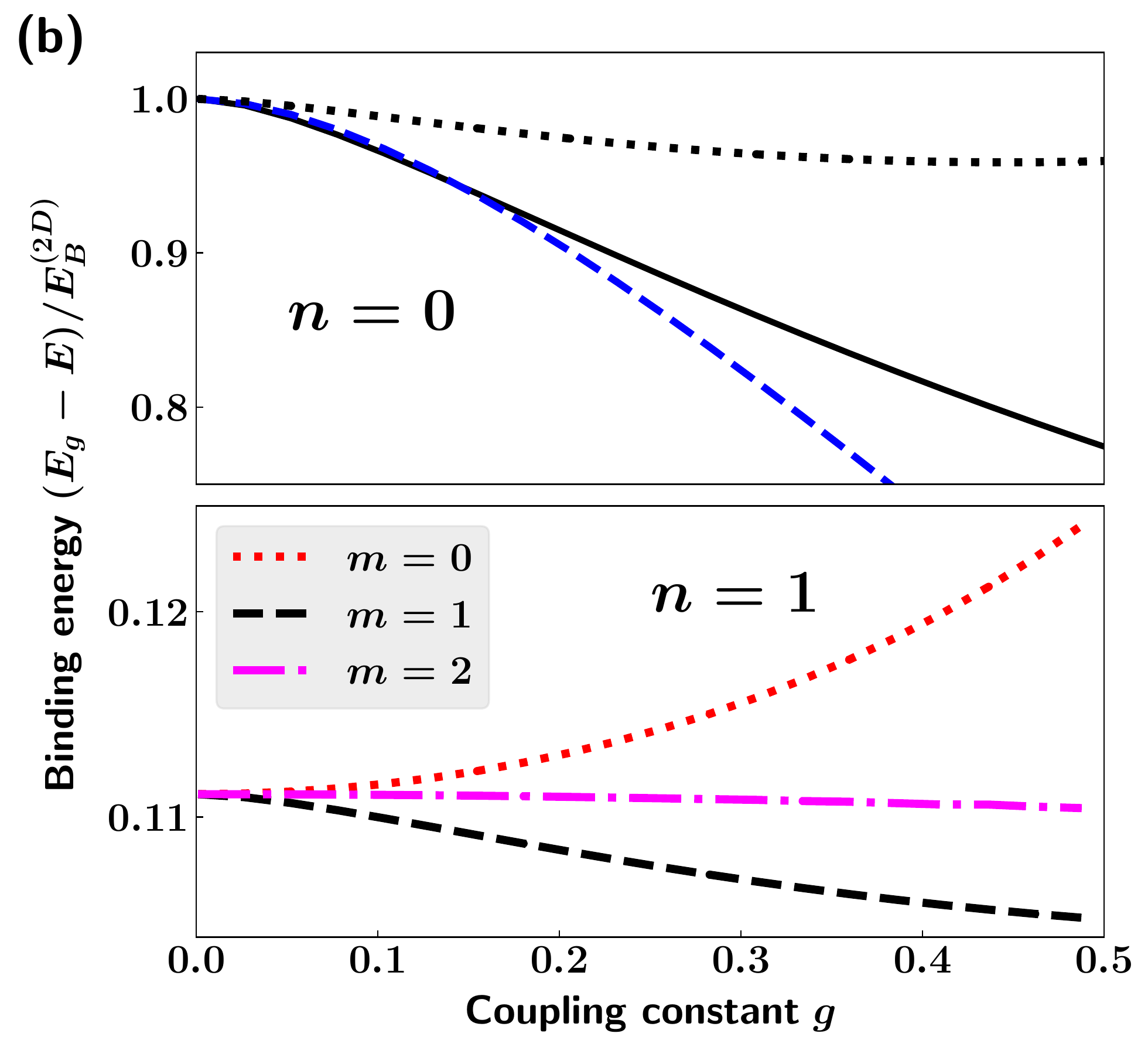}
\caption{Exciton binding energy calculated in the model with the 2D Coulomb potential (\ref{V_q}) for the lowest exciton states originating from the levels $n=0$
and $n=1$. (a)~Solid line shows the binding energy of the state ($n=0,m=1$),  the yellow dashed line is 
the non-relativistic limit~\eqref{E_B_wbg}, and the blue dashed lines are the approximation~\eqref{bind_en_approx} with $\gamma=1$.
(b) The same dependencies rescaled in units of $E_B^{(2D)}$. The dotted line for $n=0$ is a result of solution of the scalar equation for ${\cal C}_{++}$ decoupled from ${\cal C}^{+}$ and ${\cal C}_{--}$ (scalar relativistic simplification).
}
\label{fig:bind_en}
\end{figure}

For the stationary Schr\"odinger equation (\ref{mass}) the bound state energy levels are of the form, e.g. \cite{2Dexc},
\begin{equation} \label{2Dlevels}
E - E_g = - \frac{E_B^{(2D)}}{(2n + 1)^2}~~(n = 0,1,2\dots)\:,
\end{equation}
where the binding energy of the ground exciton state is~\cite{Silin,Weak_int}
\begin{equation} \label{E_B_wbg}
E_B^{(2D)} = {g^2\over 2} E_g\:.
\end{equation}
The ground state level $n = 0$ is nondegenerate and has zero angular momentum component $\ell = 0$ (or $m = 1$) while the first excited level is triple-degenerate with ${\ell = 0, \pm 1}$ (or $m = 0, 1, 2$). 
Figure~\ref{fig:bind_en} shows the ratio between the binding energy $E_g - E$ and (a) the band gap or (b) the 2D Rydberg (\ref{E_B_wbg}). 

With increasing the interaction strength (\ref{ge2}) one should take into account the nonparabolicity of the electron energy dispersion and the Coulomb-scattering induced mixing of the coefficients $C_{\lambda_e \lambda_h}$ in Eq.~(\ref{C_syst}).
In a simplified approach one may switch in the relativistic dispersion (\ref{relativ}) but neglect the mixing and retain only the coefficient $C_{++}$ in the referred equations. This means the replacement in the scalar equation (\ref{mass}) for $C_{++}({\bm k})$ the kinetic energy $\hbar^2 k^2/2 \mu$ by $2\epsilon_k - E_g$ and the Fourier image $V(|{\bm k} - {\bm k}'|)$ by the Coulomb matrix element ${\cal J}_{++,++}({\bm k} \leftarrow {\bm k}')$ in Eq.~(\ref{scattexc}). In the following we refer to this approach as to \emph{the scalar relativistic simplification}.

For the exact solution of the relativistic equation~(\ref{syst_3_eq}), it is convenient to introduce the dimensionless positive variables
\begin{equation} \label{Xepsilon}
Q = \frac{1}{g} \frac{2 \hbar v_0 k}{E_g}\:, \quad \epsilon = \frac{E_g - E}{E_B^{(2D)}} = \frac{2}{g^2}\frac{E_g - E}{E_g}
\end{equation} 
and the 2D vectors ${\bm Q}, {\bm Q}'$ determined by the absolute values $Q,Q'$ and the azimuth angles $\varphi, \varphi'$. Then dividing the left- and right-hand sides of Eq.~(\ref{syst_3_eq}) by $E_g$ we obtain in the new variables
\begin{eqnarray} \label{syst_3_eq2}
&& \left( 1 - \frac{2 \epsilon}{g^2} \right) \bm{\mathcal C}(Q,\varphi) =  \bm{\mathcal H}_0(Q)  \bm{\mathcal C}(Q,\varphi)\\
&&- \frac{g^2}{2} \sum_{{\bm Q}'} \frac{\sum\limits_{l=0, \pm 1} {\bm F}_l(gQ, gQ')\text{e}^{il(\varphi'-\varphi)}}{ \sqrt{ Q^2 + Q^{\prime 2} - 2 QQ' \cos{ ( \varphi' -\varphi)}}}\ \bm{\mathcal C}(Q',\varphi')\:. \nonumber
\end{eqnarray} 
Here the diagonal matrix $\bm{\mathcal H}_0(Q)$ and the matrix ${\bm F}_l(gQ, gQ') = S_l(gQ) S^T_l(gQ')$ are obtained from those in Eq.~(\ref{syst_3_eq}) by replacing $2 \epsilon_k$ to  $\sqrt{1 + (gQ)^2}$ and 
\[
T_{\pm}(k) \to \sqrt{\frac12 \left( 1 \pm \frac{1}{\sqrt{1 + (gQ)^2}}\right)}\:.
\]
We see that the modified equation is controlled only by one parameter, the strength $g$.

In order to solve the vector equation~\eqref{syst_3_eq2} for  the angular harmonics $\bm{\mathcal C}({\bm Q}) = \bm{\mathcal C}(Q) {\rm e}^{{\rm i} m \varphi}$ we use a modified version of the Gauss--Legendre quadrature method~\cite{Chuang_par} which avoids the singularity of the Coulomb potential at $\bm Q = \bm Q'$ in Eq.~\eqref{syst_3_eq2}. 
We introduce the variable $x$ via ${Q = \tan{(\pi x/2)}}$ and use the quadrature method with mesh points $x_{i}$ and weights $w_{i}$ for $i = 1\ldots N$.
The integration over $Q$ is replaced by a Riemann summation over $i$ via $dQ \to w_i (dQ/dx)_i$.
In the numerical calculation we take $N = 250$ and check that a further increase in $N$ does not lead to visible changes of the curves in Fig.~\ref{fig:bind_en}.

Figure~\ref{fig:bind_en} shows the $g$ dependence of the four lowest exciton levels originating from the low-$g$ levels $n=0, m=1$ and $n=1~(m=0,1,2)$. In accordance with Eq.~(\ref{2Dlevels}), at small values of $g$ the ratio $(E_g - E)/E^{(2D)}_B$ approaches~1 for $n=0$ and $1/9$ for $n=1$. 
The exciton binding energy increases with $g$
 in units of the band gap $E_g$ but decreases in units of $E^{(2D)}_B \propto g^2$. In Appendix~\ref{1_over_Eg_perturb} we find the first asymptotic correction to the binding energy in the regime $g \to 0$. The result is 
\begin{equation}
\label{bind_en_approx}
E_g -E \approx E_B^{(2D)}\left( 1 -g^2 \ln{{\gamma\over g}}\right) ,
\end{equation}
where $\gamma$ is a constant of the order of unity. The blue dashed line in Fig.~\ref{fig:bind_en} depicts this asymptotic behaviour calculated with $\gamma=1$ and demonstrating a good agreement for small $g$.
The three-fold degeneracy of the $n=1$ level is removed with the increasing value of $g$ but the sublevel splitting is small and reaches 10\% of $E^{(2D)}_B$ only at $g\approx 0.5$.

In Fig.~\ref{fig:bind_en}(b), higher panel, we compare the exciton binding energies obtained by solving Eq.~\eqref{syst_3_eq2} with the simplified scalar relativistic approach. The dotted curve calculated in this approach is in a quantitative agreement with results of Ref.~\cite{MacDonald_2015} where the one-component wavefunction was used. Figure~\ref{fig:bind_en}(b) demonstrates that this simplified calculation overestimates the binding energy by $\sim 20$\% at $g=0.5$.

\begin{figure}[h]
\includegraphics[width=0.99\linewidth]{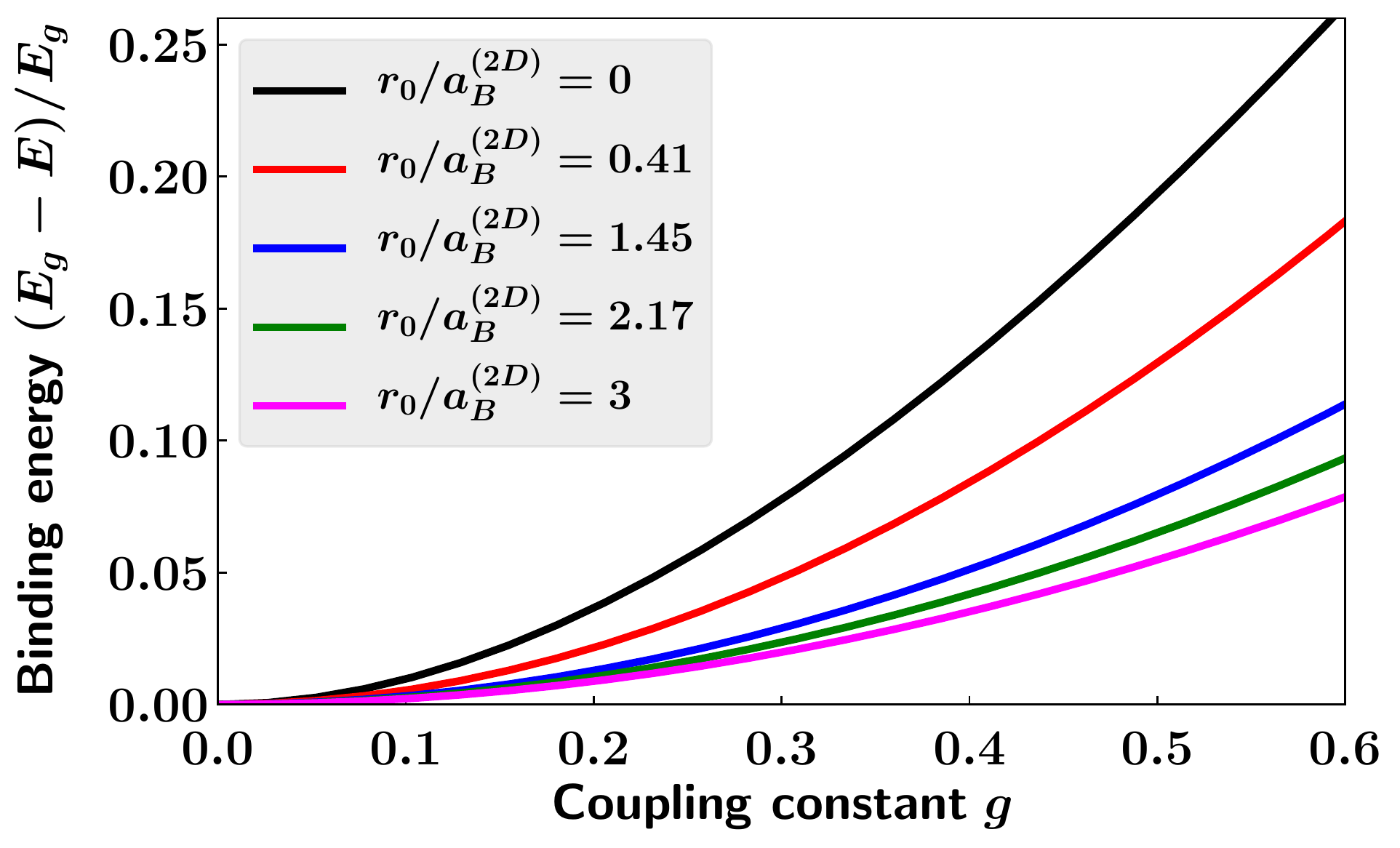}
\caption{Binding energy for the lowest exciton level $n = 0$ for different values of the screening radius. The black curve shows the case of a pure Coulomb potential.}
\label{Bind_en}
\end{figure}

Now we turn to the Rytova--Keldysh potential~(\ref{V_q2}).
Numerical solution of Eq.~\eqref{syst_3_eq2} with the screened potential yields the binding energy values shown in Fig.~\ref{Bind_en}. For comparison, the result for pure Coulomb potential ($r_0=0$) is also shown.
The binding energy grows with an increase of the coupling constant. Naturally, the screening reduces the binding energy for a given value of $g$ in comparison with the pure Coulomb potential. 

In real 2D Dirac materials, the exciton binding energy and the oscillator strength are governed by a single independent parameter, the dielectric constant $\varkappa$.
It is convenient to rewrite the Fourier-image of the screened potential \eqref{V_q2} as follows
\begin{equation}
V_{RK}(|\bm k - \bm k'|) =  - \frac{4 \pi (\hbar v_0)^2}{E_g Q_- \left(1+Q_- C/\varkappa^{2} \right)}\:.
\end{equation}
Here $Q_- = \left\vert {\bm Q} - {\bm Q}' \right\vert$, the dimensionless vector ${\bm Q}$ is introduced in Eq.~\eqref{Xepsilon}, $C$ is the TMD monolayer constant 
\begin{equation}
C = \frac{e^2 E_g l}{2 (\hbar v_0)^2} = {g_{0}^{2}E_{g}l\over 2 e^{2}}\:,
\end{equation}
and $g_0 = e^2/(\hbar v_0)$. The parameters for four dichalcogenides are given in Table~\ref{TMDPAR}.

\begin{table}[htp]
\begin{center}
\begin{tabular}{|c|c|c|c||c|c|}
\hline
&$E_g$ (eV)& $c/v_0$ & $l$ (\AA) &  $g_{0}$ & $C$ \\
\hline
MoS$_{2}$ & 1.66 & 555 & 41.47 & 4.05 & 39.25\\
\hline
MoSe$_{2}$& 1.47 & 613 & 51.71 & 4.47 & 52.85 \\
\hline
WS$_{2}$& 1.79 & 428 & 37.89 & 3.12 & 22.95\\
\hline
WSe$_{2}$& 1.60 & 466 & 45.11 & 3.4 & 29.03\\
\hline
\end{tabular}
\end{center}
\caption{Parameters of TMD monolayers from Ref.~\cite{Exc_trion_TMD} and calculated values of $g_{0}$ and $C$.}
\label{TMDPAR}
\end{table}

\begin{figure}[h]
\includegraphics[width=0.99\linewidth]{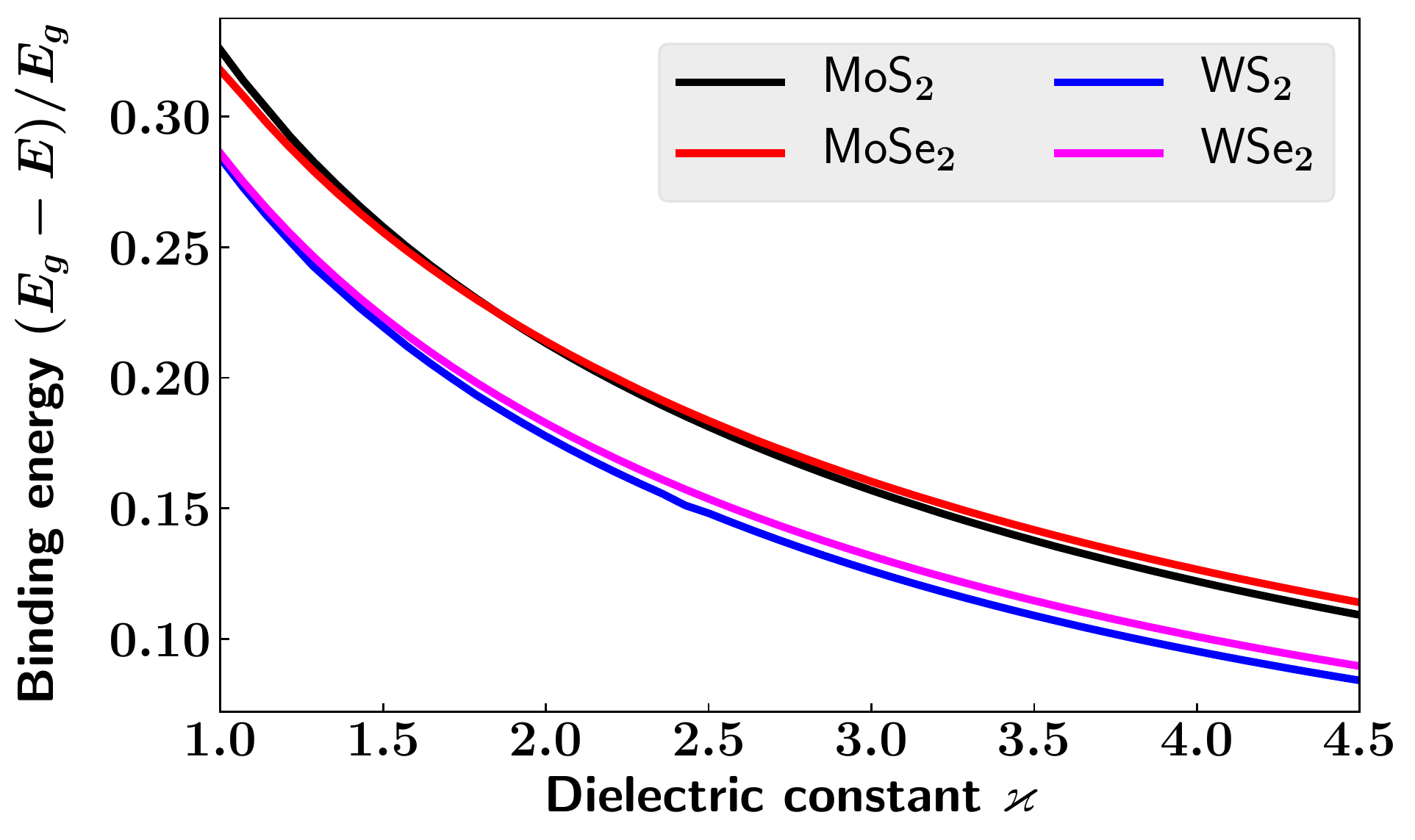}
\caption{The exciton binding energy for the lowest exciton level $n = 0$ for real TMD monolayers as a function of environmental-dependent dielectric constant.}
\label{TMD_en}
\end{figure}

Figure~\ref{TMD_en} shows the $\varkappa$-dependence of the exciton binding energy
for the four TMD monolayers. 
One can see that the binding energy is determined mostly by the transition metal rather than by the chalcogen. We have checked that the variation of the binding energy with $\varkappa$ is in a quantitative agreement with the results of Ref.~\cite{Exc_TMD_energies}.
The effect of the dielectric environment on the exciton energy is remarkable: the binding energy varies by a factor of three for all the four materials when $\varkappa$ grows from 1 to 4.5. This  decrease of the binding energy due to screening is expected, but quantitatively it is much weaker than in the non-relativistic limit for the Coulomb potential, Eq.~\eqref{E_B_wbg}, where ${E_B^{(2D)} \propto 1/\varkappa^2}$.
The latter regime is realized at larger $\varkappa$ where the coupling constant $g$ is small enough. 

\subsection{Oscillator strength}
We define the exciton oscillator strength as
\begin{equation} \label{oscstr}
\Omega \equiv |M(\sigma^+)|^2\:.
\end{equation}
In the non-relativistic limit, $E^{(2D)}_B \ll E_g$, and for the 2D Coulomb potential~(\ref{V_q}), the absolute value of the matrix element $M(\sigma^+)$ for the exciton ground state $n=0$ is given by
\begin{equation} 
\label{osc_str_wbg}
\left\vert M^{\rm nr}_0(\sigma^+;{\rm Coul})\right\vert = \frac{2|e| v_0}{\sqrt{\pi}a^{(2D)}_B} 
= \frac{g |e|E_g}{\sqrt{\pi} \hbar} 
\end{equation}
with $a^{(2D)}_B$ being the Bohr radius of a 2D exciton 
$$a^{(2D)}_B = \frac{\hbar^2 \varkappa}{2 \mu e^2} = \frac{2\hbar v_0}{g E_g}\:.$$
Estimates show that, for all the four TMD monolayers in Table~\ref{TMDPAR}, $a^{(2D)}_B/\varkappa \approx 1$~\AA.

\begin{figure}[h!]
\includegraphics[width=0.99\linewidth]{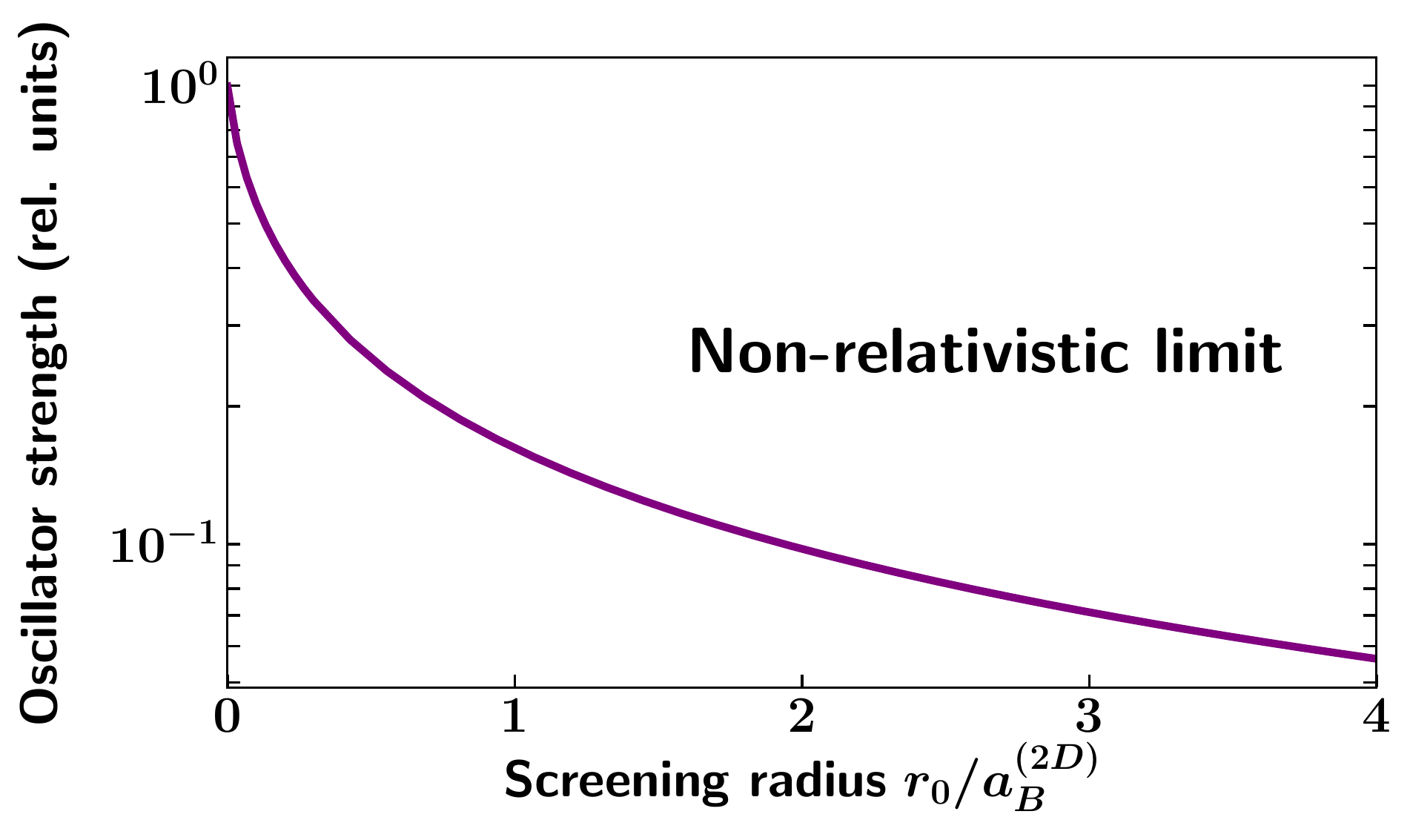}
\caption{The ratio of ground-state exciton oscillator strengths $\Omega_0^{\rm nr}({\rm RK})/\Omega_0^{\rm nr}({\rm Coul})$ calculated in the non-relativistic limit for the Rytova--Keldysh and Coulomb potentials as a function of the screening radius.
}
\label{osc_stren_r0}
\end{figure}

As compared with the Coulomb potential, the Rytova--Keldysh attraction leads to a smaller exciton binding energy and, therefore, to a weaker oscillator strength. Figure~\ref{osc_stren_r0} depicts the $r_0$-dependence of the ratio, $\Omega_0^{\rm nr}({\rm RK})/\Omega_0^{\rm nr}({\rm Coul})$, of the ground-state exciton oscillator strengths calculated for the two potentials in the non-relativistic limit $g \to 0$, $E_g \to \infty$, $a^{(2D)}_B = {\rm const.}$ The effect of screening-induced suppression of the absorption efficiency is clearly seen: for the screening radius $r_0 = 2a^{(2D)}_B$ the ratio drops by an order of magnitude.

\begin{figure}[h!]
\includegraphics[width=0.99\linewidth]{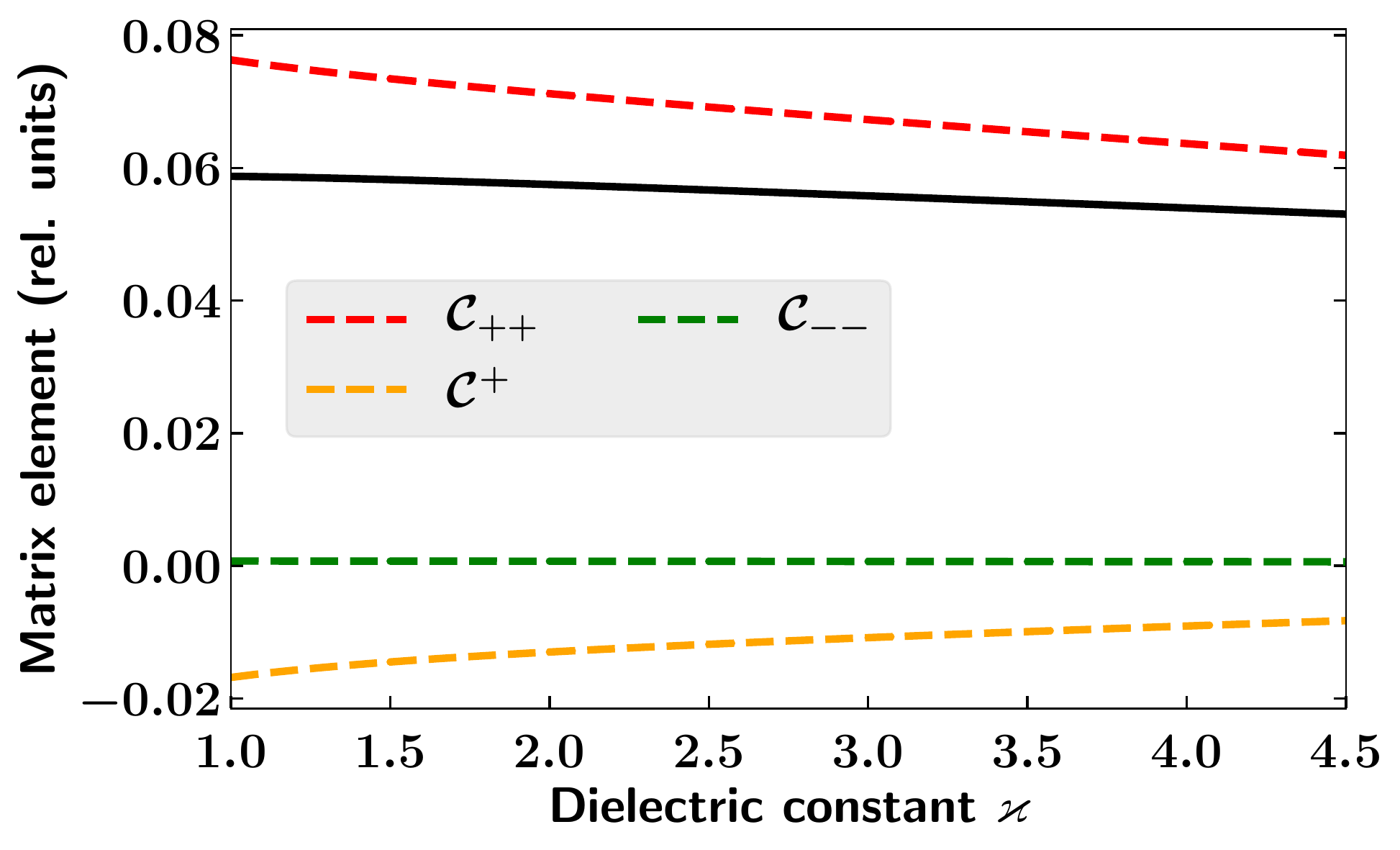}
\caption{The dependence of the optical matrix element~\eqref{osc_strength2} for the lowest exciton state ${n = 0}$ in MoS$_2$ monolayer on the dielectric constant (solid black line).  
The matrix element is given in units $M_0(\sigma^+)/\left\vert M^{\rm nr}_0(\sigma^+; {\rm Coul};{\rm vac}) \right\vert$.
Dashed lines show partial contributions to the matrix element. 
}
\label{Matr_el_kappa}
\end{figure}

Figure~\ref{Matr_el_kappa} presents the calculation of the optical matrix element $M(\sigma^+)$, Eq.~(\ref{osc_strength2}), for the exciton  ground state $n=0$ for the Rytova--Keldysh potential. 
In order to demonstrate the nonparabolicity effect we plot in this figure 
the ratio of $M_0(\sigma^+;{\rm RK})$ to $\left\vert M^{\rm nr}_0(\sigma^+; {\rm Coul};{\rm vac}) \right\vert$
calculated in the non-relativistic limit, for a suspended TMD monolayer in vacuum:
\begin{equation}
\label{M_vac}
\left\vert M^{\rm nr}_0(\sigma^+; {\rm Coul};{\rm vac}) \right\vert = \frac{|e|^3 E_g}{\sqrt{\pi} \hbar^2 v_0} .
\end{equation}
Dashed lines show the partial contributions of the three components  ${\cal C}_{++}$, ${\cal C}^+$ and ${\cal C}_{--}$  in 
Eq.~\eqref{R}. 
While the contribution from ${\cal C}_{--}$ is negligible, the term due to ${\cal C}^+$ 
is negative and its absolute value is $\sim 25$~\% of the contribution from the component ${\cal C}_{++}$.
Thus, the optical absorption efficiency is smaller than the value obtained in the scalar relativistic simple model due to the admixture of the states $|e, +, {\bm k}; h, -, -{\bm k} \rangle$ and $|e, -, {\bm k}; h, +, -{\bm k} \rangle$ to the exciton wavefunction~(\ref{Clelh}). 

In Ref.~\cite{Exc_trion_TMD} the exciton wavefunction is found in the real space and the equation for the exciton oscillator strength contains only the squared wavefunction $|\phi^{e,h,j}_{c,v}(0)|^2$  at ${\bm \rho} = 0$. In our notations the function $\phi^{e,h,j}_{c,v}({\bm \rho})$ coincides with
\[
\psi_{++}({\bm \rho}) = \sum_{\bm k} {\rm e}^{{\rm i} ({\bm k}{\bm \rho} + \varphi_{\bm k})} C_{++}({\bm k})\:.
\]
Our calculation, Fig.~\ref{Matr_el_kappa},  shows that the sum
\[
- \sqrt{2} e v_0 \sum\limits_{\bm k} {\rm e}^{{\rm i}\varphi_{\bm k}} T_+ T_- C^{+ *}({\bm k})
\]
makes a remarkable contribution
which is missed in the equation~(26) of Ref.~\cite{Exc_trion_TMD}.

\begin{figure}[h]
\includegraphics[width=0.99\linewidth]{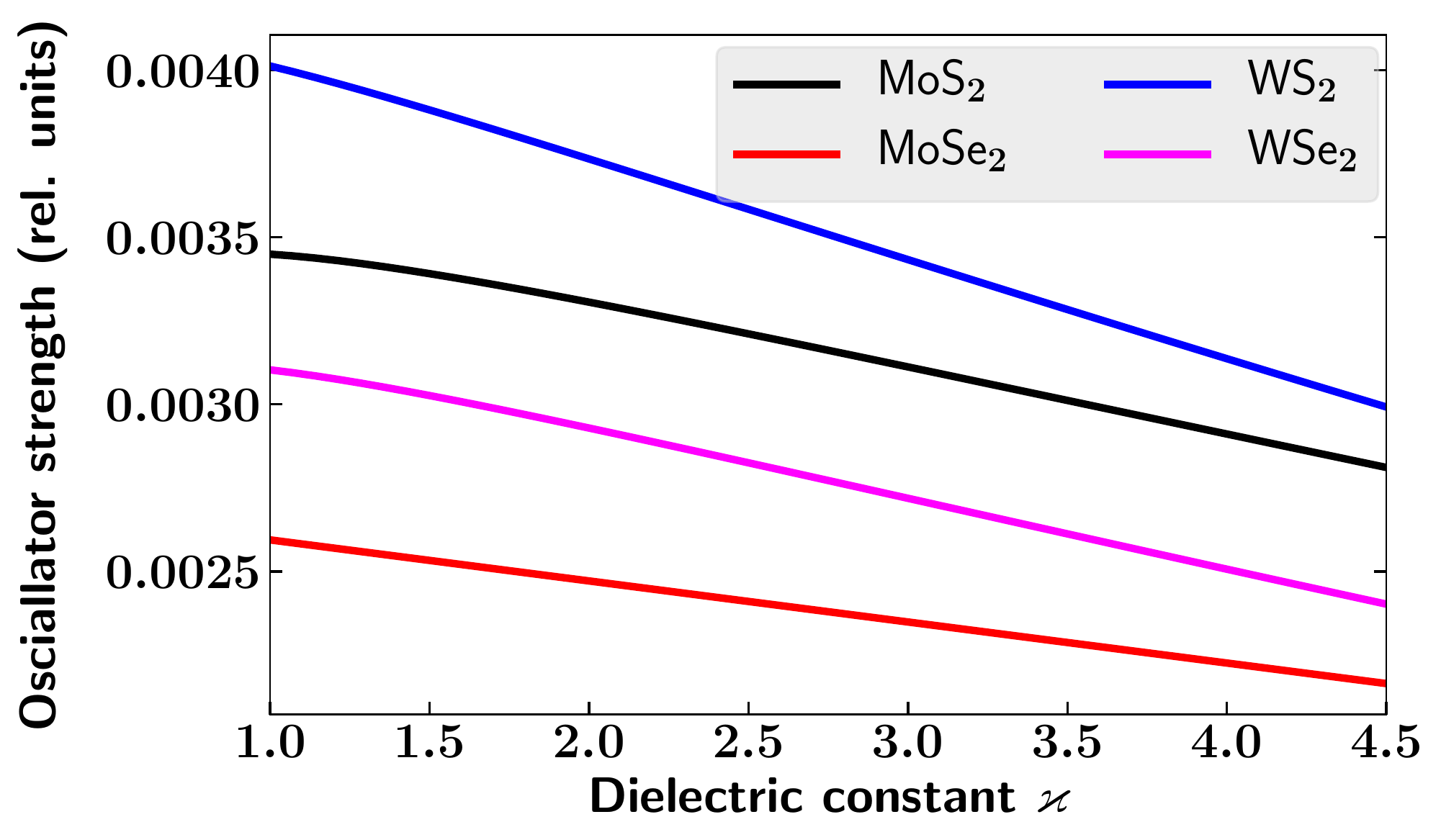}
\caption{The ratio of the oscillator strengths $\Omega_0({\rm RK})/\Omega^{\rm nr}_0({\rm Coul}; {\rm vac})$ for the lowest exciton level $n = 0$ for the TMD monolayers as a function of environmental-dependent dielectric constant $\varkappa$.
}
\label{TMD_osc}
\end{figure}

Figure~\ref{TMD_osc} shows the $\varkappa$-dependence of the oscillator strength 
for the four TMD monolayers.
The oscillator strength decreases slower
as compared to the non-relativistic limit for the Coulomb potential, Eq.~\eqref{osc_str_wbg}, which yields $\Omega^{\rm nr}_0({\rm Coul}) \propto 1/\varkappa^2$. 
This limit is achieved at the higher $\varkappa$ and the smaller coupling constant $g$.
Thus,  due to the nonparabolicity effects, 
the strength of the exciton absorption peak is
less sensitive to the dielectric constant $\varkappa$ than in the limit of parabolic free-carrier dispersion, in particular, due to a multicomponent nature of the excitonic wavefunction.

\section{Concluding remarks}
\label{Conc}

Beginning from 1960-s it has been recognized that, if the band gap $E_g$ of an intrinsic semiconductor is smaller than the exciton binding energy $E_B$, the crystal becomes unstable,
and a new phase, ``excitonic insulator'', 
emerges~\cite{Kopaev,Cloizeaux,Maksimov,Kohn}. A similar transition, the ``excitonic collapse'',  has been recently analyzed in conjunction with the TMD 2D crystals \cite{Rodin}. The existing TMD monolayers, particularly those listed in Table~\ref{TMDPAR}, are stable semiconductors. They  are nonetheless characterized by large values of the binding energy-to-gap ratio $E_B/E_g$. The studies \cite{Trushin_2016,Exc_trion_TMD,Exc_TMD_energies} show that in semiconductors with increasing the ratio $E_B/E_g$, before the many-body effects become important, the structure of the exciton wavefunction is strongly modified, Eq.~(\ref{Clelh}), and acquires new features. In this work, we have demonstrated an importance of the four-component structure of the exciton wavefunction for the description of resonant optical properties of TMD monolayers. Both the Coulomb and Rytova--Keldysh potentials have been used for the calculation of the exciton binding energy and oscillator strength.

\acknowledgments

The financial support of the Russian Science Foundation (Project~17-12-01265) is acknowledged.
The work of N.V.L. and L.E.G. was supported by the Foundation for the Advancement of Theoretical Physics and Mathematics ``BASIS''.

\appendix

\section{Weak Coulomb interaction: Expansion in powers of $g$}
\label{1_over_Eg_perturb}
Taking the electron and hole effective Hamiltonians in the form (\ref{HamiltK}) and (\ref{Hamilth})  we can explicitly rewrite Eq.~(\ref{HeHh}) as a set of four equations
\begin{subequations}
\begin{align}
\label{eq_syst1}
&(E_g + V)\psi_{++} + v_0p_-\psi_{+-} + v_0 p_-\psi_{-+} = E\psi_{++}\:,\\
\label{eq_syst2}
& v_0p_+\psi_{++} + V\psi_{+-} + v_0p_-\psi_{--}= E\psi_{+-}\:, \\
\label{eq_syst3}
&v_0p_+\psi_{++} + V\psi_{-+} + v_0p_-\psi_{--}= E\psi_{-+}\:, \\
&v_0p_+(\psi_{+-}+\psi_{-+}) + (V- E_g) \psi_{--} = E\psi_{--}\:.
\end{align}
\end{subequations}
Here $p_{\pm} = - {\rm i} \hbar( \partial/\partial_x \pm {\rm i} \partial/\partial_y)$ and $V = - e^2/(\varkappa \rho)$.
We calculate a correction to the binding energy $E^{(2D)}_B$ obtained in the effective mass theory, see Eq.~(\ref{E_B_wbg}).  

Introducing $\varepsilon=E-E_g$ and assuming $|\varepsilon| \ll E_g$, we obtain in the first order in $v_0 p/E_g$
\begin{equation}
\psi_{+-} = \psi_{-+} = {v_0\over E_g} p_+ \psi_{++}, \qquad \psi_{--}=0\:.
\end{equation}
Substituting  $\psi_{+-}, \psi_{-+}$ into Eq.~\eqref{eq_syst1} we obtain an uncoupled equation for $\psi_{++}$
\begin{equation}
\label{H_Coul}
\left(V + {2v_0^2\over E_g}p^2 \right)\psi_{++} = \varepsilon\psi_{++} \: .
\end{equation}
This is the equation for 2D Coulomb problem with the reduced effective mass $\mu=E_g/(4v_0^2)$. For the exciton ground state, the energy $\varepsilon$ equals $- E^{(2D)}_B$ and the envelope is given by
\begin{equation}
\label{2D_H_atom}
\psi_0({\bm \rho}) =\sqrt{2\over \pi} \frac{\exp{\left(-\rho/a^{(2D)}_B\right)}}{a^{(2D)}_B}\:.
\end{equation}

Now we turn to a correction of the order  $(v_0 p/E_g)^2$. 
To this order  $\psi_{--}$ becomes nonzero and is approximated by
\begin{equation}
\psi_{--}={v_0\over 2E_g}p_+(\psi_{+-}+\psi_{-+}) \: .
\end{equation}
Substituting $\psi_{--}$ into Eqs.~\eqref{eq_syst2},~\eqref{eq_syst3} we find with the second-order accuracy 
\begin{equation}
\psi_{+-}+ \psi_{-+}={2v_0\over E_g}\left(1 - {\varepsilon-V\over E_g} +{v_0^2\over E_g^2}p^2\right)p_+\psi_{++} \: .
\end{equation}
The substitution of this sum into Eq.~\eqref{eq_syst1} yields a corrected equation for the function $\psi_{++}$. In analogy with the three-dimensional Dirac problem \cite{LL_4}, we introduce, instead of $\psi_{++}$, the function
\begin{equation}
\psi_\text{Shr} = \left(1 +{v_0^2\over E_g^2} p^2  \right)\psi_{++}\:.
\end{equation}
It satisfies the following Schr\"odinger equation
\begin{equation}
({\cal H}_0+U)\psi_\text{Shr} = \varepsilon \psi_\text{Shr} \: ,
\end{equation}
where ${\cal H}_0$ is the Hamiltonian of 2D Coulomb problem~\eqref{H_Coul} and the perturbation has the form
\begin{equation}
U = - \frac{2v_0^4}{E_g^3} p^4  + \left( \frac{\hbar v_0}{E_g} \right)^2 {\bm \nabla}^2 V
+ \frac{2 \hbar v^2_0}{E^2_g} \left[ {\bm \nabla} V \times {\bm p} \right]_z \:,
\end{equation}
with ${\bm \nabla}^2$ being the 2D Laplace operator $\partial_{x}^2+\partial_{y}^2$. 

The correction to the binding energy is given by the average $\langle U \rangle = \int \psi_0({\bm \rho})U \psi_0({\bm \rho}) d {\bm \rho}$ which can be reduced to
\[
\langle U \rangle  = { 2(\hbar v_0)^4 \over E_g^3 \left(a^{(2D)}_B\right)^2} \left[ \left<\rho^{-2}\right> + \frac{2}{ a^{(2D)}_B}\left<\rho^{-1}\right>   - \frac{1}{ \left(a^{(2D)}_B\right)^2 } \right].
\]
The first term is singular at $\rho=0$, and the contributions from the other terms can be neglected. 
Assuming its integration to start from  $\rho_\text{min}=\gamma^{-1}(\hbar v_0/E_g) ={g\over 2\gamma} a^{(2D)}_B$, where $\gamma\sim 1$ and therefore $\rho_\text{min} \ll a^{(2D)}_B$, we obtain
\begin{equation}
\langle U \rangle  = {8(\hbar v_0)^4 \over E_g^3 \left(a^{(2D)}_B\right)^4}\left(-\ln{2\rho_\text{min}\over a^{(2D)}_B}\right) =  E_B^{(2D)} g^2\ln{\gamma\over g}.
\end{equation}
This yields Eq.~\eqref{bind_en_approx} of the main text.

\end{document}